\documentclass[final,5p,times,twocolumn]{elsarticle}
\usepackage{lineno,hyperref,amsmath,amsthm,amssymb,amsfonts,ragged2e,color,subfig}
\journal{``The European Physical Journal D"}
\begin{document}
\begin{frontmatter}
\title{Ion-acoustic shock waves in magnetized pair-ion plasma}
\author{T. Yeashna$^{*,1}$, R.K. Shikha$^{**,1}$,  N.A. Chowdhury$^{***,2}$, A. Mannan$^{\dag,1,3}$, S. Sultana$^{\ddag,1}$, and A.A. Mamun$^{\S,1}$}
\address{$^{1}$Department of Physics, Jahangirnagar University, Savar, Dhaka-1342, Bangladesh\\
$^2$ Plasma Physics Division, Atomic Energy Centre, Dhaka-1000, Bangladesh\\
$^3$ Institut f\"{u}r Mathematik, Martin Luther Universit\"{a}t Halle-Wittenberg, Halle, Germany\\
e-mail: $^*$yeashna147phy@gmail.com, $^{**}$shikha261phy@gmail.com, $^{***}$nurealam1743phy@gmail.com,\\
$^{\dag}$abdulmannan@juniv.edu, $^{\ddag}$ssultana@juniv.edu, $^{\S}$mamun\_phys@juniv.edu}
\begin{abstract}
A theoretical investigation associated with obliquely propagating ion-acoustic shock waves (IASHWs)
in a three-component magnetized plasma having inertialess non-extensive electrons, inertial
warm positive and negative ions has been performed. A Burgers equation is derived by employing the
reductive perturbation method. Our plasma model supports both positive and negative shock structures
under the consideration of non-extensive electrons. It is found that the positive and negative
shock wave potentials increase with the oblique angle ($\delta$) which arises due to the external magnetic field.
It is also observed that the magnitude of the amplitude of positive and negative shock waves is not
effected by the variation of the ion kinematic viscosity but the steepness of the  positive and negative shock waves decreases with
ion kinematic viscosity. The implications of our findings in space and laboratory plasmas are briefly discussed.
\end{abstract}
\begin{keyword}
Pair-ion \sep  Magnetized plasma \sep Ion-acoustic waves \sep  Perturbation methods  \sep  Shock waves.
\end{keyword}
\end{frontmatter}
\section{Introduction}
\label{1sec:Introduction}
The pair-ion (PI) plasma can be observed in astrophysical environments such as
upper regions of Titan's atmosphere \cite{Coates2007,Massey1976,Sabry2009,Abdelwahed2016,Misra2009,Mushtaq2012,Jannat2015,El-Labany2020},
cometary comae \cite{Chaizy1991}, ($H^+$, $O_2^-$) and ($H^+$, $H^-$) plasmas in the D and F-regions of Earth's
ionosphere \cite{Massey1976,Sabry2009,Abdelwahed2016,Misra2009,Mushtaq2012,Jannat2015}, and also
in the laboratory experiments namely, ($Ar^+$, $F^-$) plasma \cite{Nakamura1984}, ($K^+$, $SF_6^-$) plasma \cite{Song1991,Sato1994},
neutral beam sources \cite{Bacal1979}, plasma processing reactors \cite{Gottscho1986},  ($Ar^+$, $SF_6^-$) plasma \cite{Wong1975,Nakamura1997,Cooney1991,Nakamura1999},
combustion products \cite{Sheehan1988}, plasma etching \cite{Sheehan1988}, ($Xe^+$, $F^-$) plasma \cite{Ichiki2002}, ($Ar^+$, $O_2^-$) plasma,
and Fullerene ($C_{60}^+$, $C_{60}^-$) plasma \cite{Oohara2003,Hatakeyama2005,Oohara2005}, etc.
Positive ions are produced by electron impact ionization, and negative ions are produced
by attachment of the low energy electrons. A number of authors studied the nonlinear
electrostatic structures in PI plasma \cite{Sabry2009,Abdelwahed2016,Misra2009,Mushtaq2012,Jannat2015,El-Labany2020}.

Highly energetic particles have been observed in the galaxy clusters \cite{Hansen2005},
the Earth’s bow-shock \cite{Asbridge1968}, in the upper ionosphere of Mars \cite{Lundlin1989},
in the vicinity of the Moon \cite{Futaana2003}, and in the magnetospheres of Jupiter and Saturn \cite{Krimigis1983}. Maxwellian velocity distribution
demonstrating the thermally equilibrium state of particles is not appropriate for explaining the dynamics of
these highly energetic particles. Renyi \cite{Renyi1955} first introduced the non-extensive $q$-distribution for explaining the
dynamics of these highly energetic particles, and further development of $q$-distribution
has been demonstrated by Tsallis \cite{Tsallis1988}. The parameter $q$ in the non-extensive $q$-distribution
describes the deviation of the plasma particles from the thermally equilibrium state. It should be noted that
$q=1$ refers to Maxwellian, and $q < 1$ ($q > 1$) refers to super-extensivity (sub-extensivity).
Jannat \textit{et al.} \cite{Jannat2015} investigated the ion-acoustic (IA) shock waves (IASHWs)
in PI plasma in the presence of non-extensive electrons, and observed that the height
of the positive potential decreases (increases) with positive (negative) ion mass.
Hussain \textit{et al.} \cite{Hussain2013} considered inertial
PI and inertialess non-extensive electrons and investigated IASHWs by considering kinematic viscosities
of both positive and negative ion species, and observed that the amplitude of the positive IASHWs decreases with $q$.
Tribeche \textit{et al.} \cite{Tribeche2010} studied IA solitary waves in a two-component
plasma, and found that the magnitude of the amplitude of positive and negative solitary structures increases
with super-extensive and sub-extensive electrons.

A plasma medium having considerable dissipative properties dictates the formation of shock
structures \cite{Hafez2017,Abdelwahed2016b,Hossen2017aa}. The Landau damping, kinematic viscosity
among the plasma species, and the collision between plasma species are the major causes of the dissipation
which is mainly responsible for the formation of shock structures in the plasma medium \cite{Hafez2017,Abdelwahed2016b,Hossen2017aa}.
The presence of kinematic viscosity plays a pivotal role in generating nonlinear waves \cite{Hafez2017,Abdelwahed2016b,Hossen2017aa}.
Hafez \textit{et al.} \cite{Hafez2017} observed that the steepness of the IASHWs decreases with the increase
of ion kinematic viscosity but the amplitude of IASHWs remains unchanged. Abdelwahed \textit{et al.} \cite{Abdelwahed2016b}
investigated IASHWs in PI plasma and reported that the kinematic viscosity coefficient
of the ion reduces the steepness of the IASHWs.

The external magnetic field is to be considered to change the dynamics of the plasma medium, and associated
electrostatic nonlinear structures. Hossen \textit{et al.} \cite{Hossen2017aa} studied the
electrostatic shock structures in magnetized dusty plasma, and found that the magnitude of the positive and negative
shock profiles increases with the oblique angle ($\delta$) which arises due to the external magnetic field.
El-Labany \textit{et al.} \cite{El-Labany2020} considered a three-component
plasma model having inertial PI and inertialess non-extensive electrons, and investigated IASHWs, and found that
the amplitude of the positive shock profile decreases with $q$. To the best knowledge of the authors, no attempt
has been made to study the IASHWs in a three-component magnetized plasma by considering kinematic viscosities
of both inertial warm positive and negative ion species, and inertialess non-extensive electrons. The aim of the present
investigation is, therefore, to derive Burgers' equation and investigate IASHWs in a three-component magnetized PI plasma,
and to observe the effects of various plasma parameters on the configuration of IASHWs.

The outline of the paper is as follows: The basic equations are displayed in section \ref{1sec:Governing equations}.
The Burgers equation has been derived in section \ref{1sec:Derivation of the Burgers equation}.
Results and discussion are reported in section \ref{1sec:Results and discussion}.
A brief conclusion is provided in section \ref{1sec:Conclusion}.
\section{Governing equations}
\label{1sec:Governing equations}
We consider a magnetized plasma system comprising inertial negatively and positively charged warm ions,
and inertialess electrons featuring $q$-distribution.
An external magnetic field $\mathbf{B}_0$ has been considered in the system directed along
the $z$-axis defining $\mathbf{B}_0 = B_0\hat{z}$, where $B_0$ and $\hat{z}$ are the strength
of the external magnetic field and unit vector directed along the $z$-axis, respectively.
The dynamics of the magnetized PI plasma system is governed by the following
set of equations \cite{Atteyaa2018,Adhikary2012,Dev2018,Dev2016,Dev2014,Deka2018,Sahu2014}
\begin{eqnarray}
&&\hspace*{-1.3cm}\frac{\partial \tilde{n}_{+}}{\partial \tilde{t}}+\acute{\nabla}\cdot(\tilde{n}_+ \tilde{u}_+)=0,
\label{1eq:1}\\
&&\hspace*{-1.3cm}\frac{\partial \tilde{u}_+}{\partial\tilde{t}}+(\tilde{u}_+\cdot\acute{\nabla})\tilde{u}_{+}=-\frac{Z_+e}{m_+}\acute{\nabla}\tilde{\psi} +\frac{Z_+eB_0}{m_+}(\tilde{u}_{+}\times\hat{z})
\nonumber\\
&&\hspace*{1.5cm}-\frac{1}{m_+n_+}\acute{\nabla} P_+ +\tilde{\eta}_+\acute{\nabla}^2\tilde{u}_+,
\label{1eq:2}\\
&&\hspace*{-1.3cm}\frac{\partial\tilde{n}_-}{\partial \tilde{t}}+\acute{\nabla}\cdot(\tilde{n}_-\tilde{u}_-)=0,
\label{1eq:3}\\
&&\hspace*{-1.3cm}\frac{\partial\tilde{u}_-}{\partial \tilde{t}}+(\tilde{u}_-\cdot\acute{\nabla})\tilde{u}_{-}=\frac{Z_-e}{m_-}\acute{\nabla}\tilde{\psi} -\frac{Z_-eB_0}{m_-}(\tilde{u}_-\times\hat{z})
\nonumber\\
&&\hspace*{1.5cm}-\frac{1}{m_-\tilde{n}_-}\acute{\nabla} P_- +\tilde{\eta}_-\acute{\nabla}^2\tilde{u}_-,
\label{1eq:4}\\
&&\hspace*{-1.3cm}\acute{\nabla}^2\tilde{\psi}=4\pi e[\tilde{n}_e+Z_-\tilde{n}_--Z_+\tilde{n}_+],
\label{1eq:5}\
\end{eqnarray}
where $\tilde{n}_+$ ($\tilde{n}_-$) is the positive (negative) ion number density, $m_+$ ($m_-$) is
the positive (negative) ion mass, $Z_+$ ($Z_-$) is the charge state of the positive (negative) ion,
$e$ being the magnitude of electron charge, $\tilde{u}_+$ ($\tilde{u}_-$) is the positive (negative)
ion fluid velocity, $\tilde{\eta}_+$ ($\tilde{\eta}_-$) is the kinematic viscosity of the positive (negative)
ion, $P_+$ ($P_-$) is the pressure of positive (negative) ion, and $\tilde{\psi}$ represents the electrostatic wave potential.
Now, we  are introducing normalized variables, namely, $n_+\rightarrow\tilde{n}_+/n_{+0}$, $n_-\rightarrow\tilde{n}_-/n_{-0}$, and $n_e\rightarrow\tilde{n}_e/n_{e0}$,
where  $n_{-0}$, $n_{+0}$, and $n_{e0}$ are the equilibrium number densities of the negative ions, positive ions, and electrons, respectively;
$u_+\rightarrow\tilde{u}_+/C_-$, $u_-\rightarrow\tilde{u}_-/C_-$ [where $C_-=(Z_-k_BT_e/m_-)^{1/2}$, $k_B$ being the Boltzmann constant, and $T_e$ being temperature of the electron]; $\psi\rightarrow\tilde{\psi}e/k_BT_e$; $t=\tilde{t}/\omega_{P_-}^{-1}$ [where $\omega_{P_-}^{-1}=(m_-/4\pi e^{2}Z_-^{2}n_{-0})^{1/2}$]; $\nabla=\acute{\nabla}/\lambda_{D}$ [where $\lambda_{D}=(k_BT_e/4\pi e^2Z_-n_{-0})^{1/2}$]. The pressure term of the positive and negative ions can be recognized as $P_{\pm}=P_{\pm0}(N_\pm/n_{\pm0})^\gamma$ with $P_{\pm0}=n_{\pm0}k_BT_\pm$ being the equilibrium
pressure of the positive (for $+0$ sign) and negative (for $-0$ sign) ions, and $T_+$ ($T_-$) being the temperature of warm positive (negative) ion, and
$\gamma=(N+2)/N$ (where $N$ is the degree of freedom and for three-dimensional case
$N=3$, then $\gamma=5/3$). For simplicity, we have considered ($\tilde{\eta}_+\approx\tilde{\eta}_-=\eta$), and $\eta$ is
normalized by $\omega_{p_-}\lambda_D^{2}$. The quasi-neutrality condition at equilibrium for our plasma model
can be written as $n_{e0} + Z_-n_{-0} \approx Z_+n_{+0}$. Equations \eqref{1eq:1}$-$\eqref{1eq:5} can be
expressed in the normalized form as \cite{Jannat2015,El-Labany2020}:
\begin{eqnarray}
&&\hspace*{-1.3cm}\frac{\partial n_+}{\partial t}+\nabla\cdot(n_+u_+)=0,
\label{1eq:6}\\
&&\hspace*{-1.3cm}\frac{\partial u_+}{\partial t}+(u_+\cdot\nabla)u_+=-\alpha_1\nabla\psi+\alpha_1\Omega_c(u_+\times\hat{z})
\nonumber\\
&&\hspace*{1.5cm}-\alpha_2\nabla n_+^{\gamma-1}+\eta\nabla^2u_+,
\label{1eq:7}\\
&&\hspace*{-1.3cm}\frac{\partial n_-}{\partial t}+\nabla\cdot(n_-u_-)=0,
\label{1eq:8}\\
&&\hspace*{-1.3cm}\frac{\partial u_-}{\partial t}+(u_-\cdot\nabla)u_-=\nabla\psi-\Omega_c(u_-\times\hat{z})
\nonumber\\
&&\hspace*{1.5cm}-\alpha_3\nabla n_-^{\gamma-1}+\eta\nabla^2u_-,
\label{1eq:9}\\
&&\hspace*{-1.3cm}\nabla^2\psi=\mu_en_e-(1+\mu_e)n_++n_-.
\label{1eq:10}\
\end{eqnarray}
Other plasma parameters are defined as $\alpha_1=Z_+m_-/Z_-m_+$,
$\alpha_2=\gamma T_+m_-/(\gamma-1)Z_-T_em_+$,
$\alpha_3=\gamma T_-/(\gamma-1)Z_-T_e$, $\mu_e = n_{e0}/Z_-n_{-0}$, and $\Omega_c=\omega_c/\omega_{p_-}$ [where $\omega_c=Z_-eB_0/m_-$].
Now, the expression for the number density of electrons following non-extensive $q$-distribution can be written as \cite{El-Labany2020}
\begin{eqnarray}
&&\hspace*{-1.3cm}n_e=\Big[1 +(q-1)\psi\Big]^{\frac{q+1}{2(q-1)}},
\label{1eq:11}\
\end{eqnarray}
where the parameter $q$ represents the non-extensive properties of electrons.
We have neglected the effect of the external magnetic field on the non-extensive electron distribution.
This is valid due to the fact that the Larmor radii of electrons is so small that as if the electrons
are flowing along the magnetic field lines of force.
Now, by substituting Eq. \eqref{1eq:11} into the Eq. \eqref{1eq:10}, and expanding up to third order in $\psi$, we get
\begin{eqnarray}
&&\hspace*{-1.3cm}\nabla^2 \psi=\mu_e+n_--(1+\mu_e)n_++\sigma_1\psi
\nonumber\\
&&\hspace*{0.0cm}+\sigma_2\psi^2+\sigma_3\psi^3+\cdot\cdot\cdot,
\label{1eq:12}\
\end{eqnarray}
where
\begin{eqnarray}
&&\hspace*{-1.3cm}\sigma_1=[\mu_e(q+1)]/2,~~~~\sigma_2=[\mu_e(q+1)(3-q)]/8,
\nonumber\\
&&\hspace*{-1.3cm}\sigma_3=[\mu_e(q+1)(3-q)(5-3q)]/48.
\nonumber\
\end{eqnarray}
We note that the terms containing $\sigma_1$, $\sigma_2$, and $\sigma_3$ are the contribution of $q$-distributed electrons.
\section{Derivation of the Burgers' equation}
\label{1sec:Derivation of the Burgers equation}
To derive the Burgers' equation for the IASHWs propagating in a magnetized PI plasma,
first we introduce the stretched co-ordinates \cite{Hossen2017aa,Washimi1966}
\begin{eqnarray}
&&\hspace*{-1.3cm}\xi=\epsilon(l_xx+l_yy+l_zz-v_p t),
\label{1eq:13}\\
&&\hspace*{-1.3cm}\tau={\epsilon}^2 t,
\label{1eq:14}\
\end{eqnarray}
where $v_p$ is the phase speed and $\epsilon$ is a smallness parameter measuring the weakness of
the dissipation ($0<\epsilon<1$). The $l_x$, $l_y$, and $l_z$ (i.e., $l_x^2+l_y^2+l_z^2=1$) are
the directional cosines of the wave vector $k$ along $x$, $y$, and $z$-axes, respectively. Then,
the dependent variables can be expressed in power series of $\epsilon$ as \cite{Hossen2017aa}
\begin{eqnarray}
&&\hspace*{-1.3cm}n_{+}=1+\epsilon n_{+}^{(1)}+\epsilon^2 n_{+}^{(2)}+\epsilon^3 n_{+}^{(3)}+\cdot\cdot\cdot,
\label{1eq:15}\\
&&\hspace*{-1.3cm}n_{-}=1+\epsilon n_{-}^{(1)}+\epsilon^2 n_{-}^{(2)}+\epsilon^3 n_{-}^{(3)}+\cdot\cdot\cdot,
\label{1eq:16}\\
&&\hspace*{-1.3cm}u_{+x,y}=\epsilon^2 u_{+x,y}^{(1)}+\epsilon^3 u_{+x,y}^{(2)}+\cdot\cdot\cdot,
\label{1eq:17}\\
&&\hspace*{-1.3cm}u_{-x,y}=\epsilon^2 u_{-x,y}^{(1)}+\epsilon^3 u_{-x,y}^{(2)}+\cdot\cdot\cdot,
\label{1eq:18}\\
&&\hspace*{-1.3cm}u_{+z}=\epsilon u_{+z}^{(1)}+\epsilon^2 u_{+z}^{(2)}+\cdot\cdot\cdot,
\label{1eq:19}\\
&&\hspace*{-1.3cm}u_{-z}=\epsilon u_{-z}^{(1)}+\epsilon^2 u_{-z}^{(2)}+\cdot\cdot\cdot,
\label{1eq:20}\\
&&\hspace*{-1.3cm}\psi=\epsilon \psi^{(1)}+\epsilon^2\psi^{(2)}+\cdot\cdot\cdot.
\label{1eq:21}\
\end{eqnarray}
Now, by substituting Eqs. \eqref{1eq:13}$-$\eqref{1eq:21} into Eqs. \eqref{1eq:6}$-$\eqref{1eq:9}, and
\eqref{1eq:12}, and collecting the terms containing $\epsilon$, the first-order equations reduce to
\begin{eqnarray}
&&\hspace*{-1.3cm} n_{+}^{(1)}=\frac{3\alpha_1l_z^2}{3v_p^2-2\alpha_2l_z^2}\psi^{(1)},
\label{1eq:22}\\
&&\hspace*{-1.3cm}u_{+z}^{(1)}=\frac{3v_p\alpha_1l_z}{3v_p^2-2\alpha_2l_z^2}\psi^{(1)},
\label{1eq:23}\\
&&\hspace*{-1.3cm}n_{-}^{(1)}=-\frac{3l_z^2}{3v_p^2-2\alpha_3l_z^2}\psi^{(1)},
\label{1eq:24}\\
&&\hspace*{-1.3cm}u_{-z}^{(1)}=-\frac{3v_pl_z}{3v_p^2-2\alpha_3l_z^2}\psi^{(1)}.
\label{1eq:25}\
\end{eqnarray}
Now, the phase speed of IASHWs can be written as
\begin{eqnarray}
&&\hspace*{-1.3cm}v_{p}\equiv v_{p+}=l_z\sqrt{{\frac{-a_1+\sqrt{a_1^2-36\sigma_1a_2}}{18\sigma_1}}},
\label{1eq:26}\\
&&\hspace*{-1.3cm}v_{p}\equiv v_{p-}=l_z\sqrt{{\frac{-a_1-\sqrt{a_1^2-36\sigma_1a_2}}{18\sigma_1}}},
\label{1eq:27}\
\end{eqnarray}
where $a_1=-9-6\alpha_2\sigma_1-6\alpha_3\sigma_1-9\alpha_1\mu_e-9\alpha_1$
and $a_2=6\alpha_2+4\alpha_2\alpha_3\sigma_1+6\alpha_1\alpha_3\mu_e+6\alpha_1\alpha_3$.
The $x$ and $y$-components of the first-order momentum equations can be manifested as
\begin{eqnarray}
&&\hspace*{-1.3cm}u_{+x}^{(1)}=-\frac{3l_yv_p^2}{\Omega_{c}(3v_p^2-2\alpha_2l_z^2)}~\frac{\partial\psi^{(1)}}{\partial\xi},
\label{1eq:28}\\
&&\hspace*{-1.3cm}u_{+y}^{(1)}=\frac{3l_xv_p^2}{\Omega_{c}(3v_p^2-2\alpha_2l_z^2)}~\frac{\partial\psi^{(1)}}{\partial\xi},
\label{1eq:29}\\
&&\hspace*{-1.3cm}u_{-x}^{(1)}=-\frac{3l_yv_p^2}{\Omega_{c}(3v_p^2-2\alpha_3l_z^2)}~\frac{\partial\psi^{(1)}}{\partial\xi},
\label{1eq:30}\\
&&\hspace*{-1.3cm} u_{-y}^{(1)}=\frac{3l_xv_p^2}{\Omega_{c}(3v_p^2-2\sigma_3l_z^2)}~\frac{\partial \psi^{(1)}}{\partial\xi}.
\label{1eq:31}\
\end{eqnarray}
Now, by taking the next higher-order terms, the equation of continuity, momentum equation, and Poisson's equation can be written as
\begin{eqnarray}
&&\hspace*{-1.3cm}\frac{\partial n_{+}^{(1)}}{\partial\tau}-v_p\frac{\partial n_{+}^{(2)}}{\partial\xi}+l_x\frac{\partial u_{+x}^{(1)}}{\partial\xi}+l_y\frac{\partial u_{+y}^{(1)}}{\partial\xi}
\nonumber\\
&&\hspace*{1.5cm}+l_z\frac{\partial u_{+z}^{(2)}}{\partial\xi}+l_z\frac{\partial}{\partial\xi}\big(n_{+}^{(1)}u_{+z}^{(1)}\big)=0,
\label{1eq:32}\\
&&\hspace*{-1.3cm}\frac{\partial u_{+z}^{(1)}}{\partial\tau}-v_p\frac{\partial u_{+z}^{(2)}}{\partial\xi}+l_zu_{+z}^{(1)}\frac{\partial u_{+z}^{(1)}}{\partial\xi}+\alpha_1l_z\frac{\partial\psi^{(2)}}{\partial\xi}
\nonumber\\
&&\hspace*{0.5cm}+\alpha_2l_z\frac{\partial }{\partial\xi}\bigg[\frac{2}{3}n_{+}^{(2)}-\frac{1}{9}(n_{+}^{(1)})^2\bigg]-\eta\frac{\partial^2u_{+z}^{(1)}}{\partial\xi^2}=0,
\label{1eq:33}\\
&&\hspace*{-1.3cm}\frac{\partial n_{-}^{(1)}}{\partial\tau}-v_p\frac{\partial n_{-}^{(2)}}{\partial\xi}+l_x\frac{\partial u_{-x}^{(1)}}{\partial\xi}+l_y\frac{\partial u_{-y}^{(1)}}{\partial\xi}
\nonumber\\
&&\hspace*{1.5cm}+l_z\frac{\partial u_{-z}^{(2)}}{\partial\xi}+l_z\frac{\partial}{\partial\xi}\big(n_{-}^{(1)}u_{-z}^{(1)}\big)=0,
\label{1eq:34}\\
&&\hspace*{-1.3cm}\frac{\partial u_{-z}^{(1)}}{\partial\tau}-v_p\frac{\partial u_{-z}^{(2)}}{\partial\xi}+l_zu_{-z}^{(1)}\frac{\partial u_{-z}^{(1)}}{\partial\xi}-l_z\frac{\partial\psi^{(2)}}{\partial\xi}
\nonumber\\
&&\hspace*{0.5cm}+\alpha_3l_z\frac{\partial }{\partial\xi}\bigg[\frac{2}{3}n_{-}^{(2)}-\frac{1}{9}(n_{-}^{(1)})^2\bigg]-\eta\frac{\partial^2u_{-z}^{(1)}}{\partial\xi^2}=0,
\label{1eq:35}\\
&&\hspace*{-1.3cm}\sigma_1\psi^{(2)}+\sigma_2{[\psi^{(1)}]}^2+n_-^{(2)}-(\mu_e+1)n_+^{(2)}=0.
\label{1eq:36}\
\end{eqnarray}
\begin{figure}
\centering
\begin{minipage}{.45\linewidth}
\includegraphics[width=\linewidth]{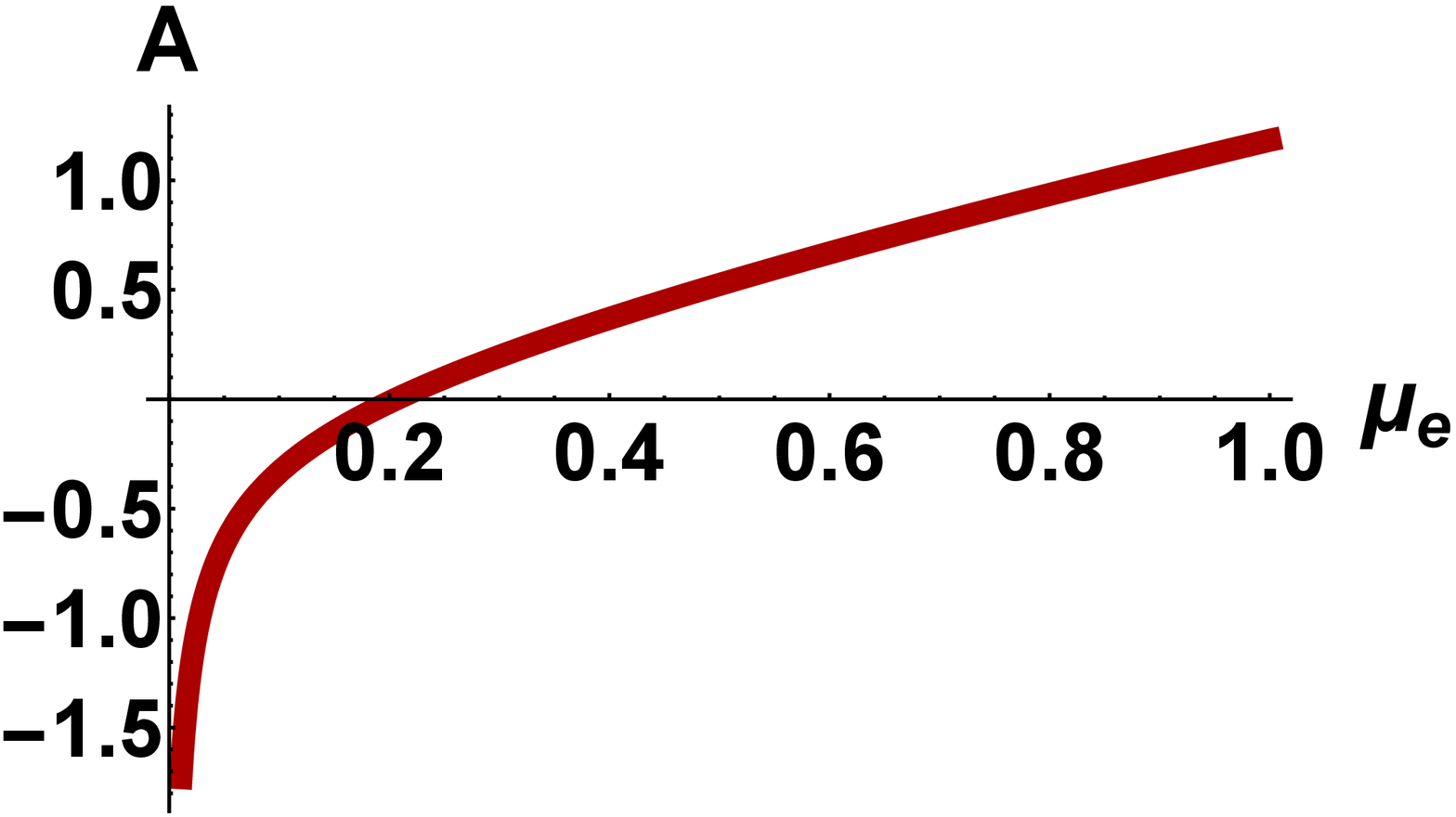}
\end{minipage}
\hspace{.05\linewidth}
\begin{minipage}{.45\linewidth}
\includegraphics[width=\linewidth]{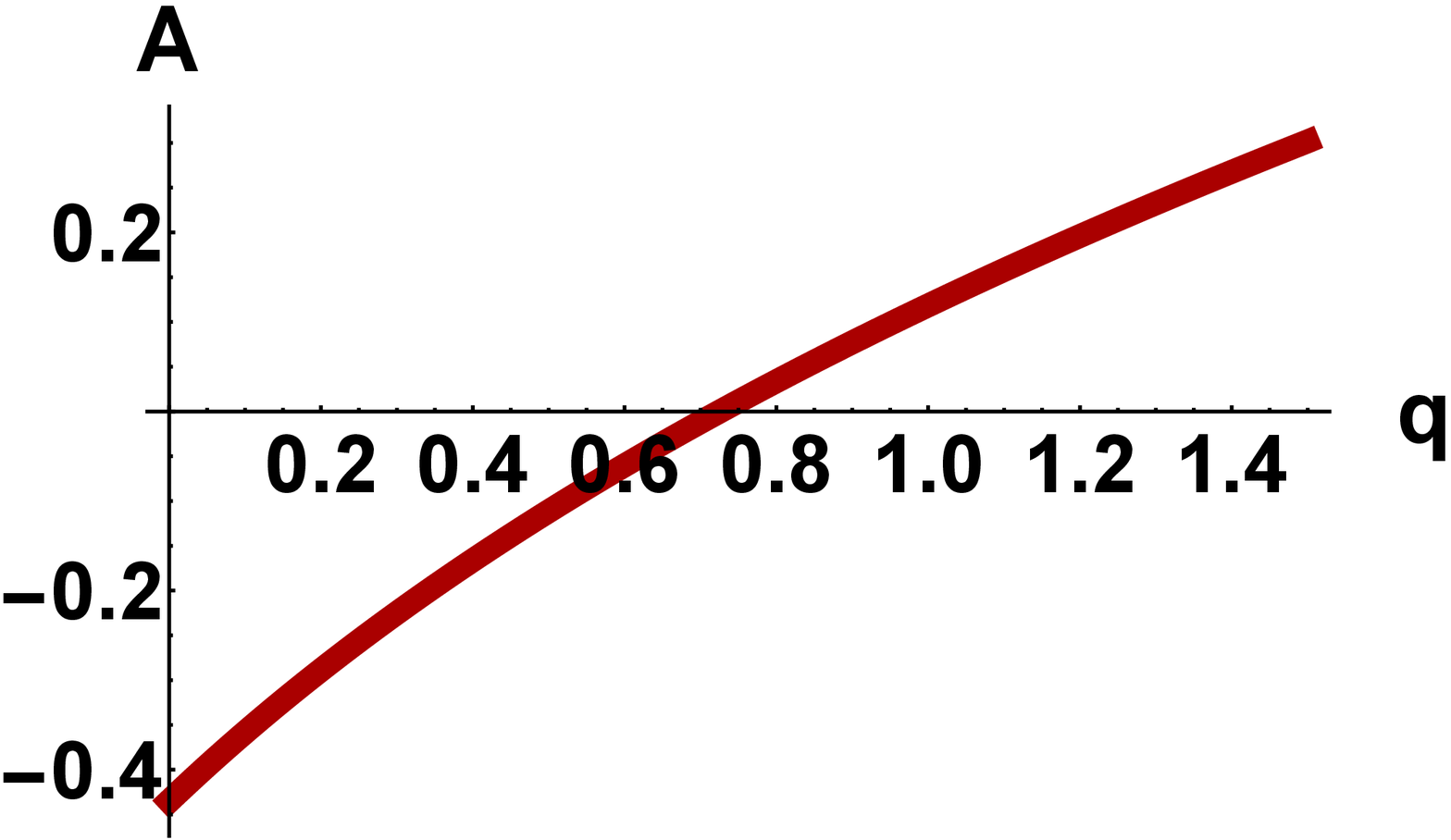}
\end{minipage}
\caption{The variation of nonlinear coefficient $A$ with $\mu_e$ when $q=1.2$ (left panel), and
the variation of nonlinear coefficient $A$ with $q$ when $\mu_e=0.3$ (right panel). Other plasma
parameters are $\alpha_1=1.5$, $\alpha_2 =0.2$, $\alpha_3=0.02$, $\delta = 30^\circ$, and $v_{p}\equiv v_{p+}$.}
\label{1Fig:F1:a:b}
\end{figure}
Finally, the next higher-order terms of Eqs. \eqref{1eq:6}$-$\eqref{1eq:9}, and \eqref{1eq:12}, with the help of
Eqs. \eqref{1eq:22}$-$\eqref{1eq:36}, can provide the Burgers equation as
\begin{eqnarray}
&&\hspace*{-1.3cm}\frac{\partial\Psi}{\partial\tau}+A\Psi\frac{\partial\Psi}{\partial\xi}=C\frac{\partial^2\Psi}{\partial\xi^2},
\label{1eq:37}\
\end{eqnarray}
where $\Psi=\psi^{(1)}$ is used for simplicity. In Eq. \eqref{1eq:37}, the nonlinear coefficient $A$ and dissipative coefficient $C$ are given by
\begin{eqnarray}
&&\hspace*{-1.3cm}A=\frac{81\alpha_1^2v_p^2s_1^3l_z^4+F_1}{18v_ps_1l_z^2s_2^3+F_2},~~~\mbox{and}~~~C = \frac{\eta}{2},
\label{1eq:38}\
\end{eqnarray}
where
\begin{eqnarray}
&&\hspace*{-1.3cm}F_1 =81\mu_e\alpha_1^2v_p^2s_1^3l_z^4-81v_p^2s_1^3l_z^4+2\mu_e\alpha_2\alpha_1^2s_1^3l_z^6
\nonumber\\
&&\hspace*{-0.5cm}+2\alpha_2\alpha_1^2s_1^3l_z^6+2\alpha_3s_1^3l_z^6-2\sigma_2s_1^3s_2^3,
\nonumber\\
&&\hspace*{-1.3cm}F_2=18\alpha_1v_ps_2l_z^2s_1^3+18\alpha_1\mu_ev_ps_2l_z^2s_1^3 ,
\nonumber\\
&&\hspace*{-1.3cm}s_1=3v_p^2-2\alpha_3l_z^2,~~~s_2 = 3v_p^2 - 2\alpha_2l_z^2 .
\nonumber\
\end{eqnarray}
Now, we look for stationary shock wave solution of this Burgers' equation  by
considering $\zeta =\xi-U_0\tau'$ and $\tau =\tau'$ (where $U_0$ is the speed of the shock waves in the reference frame).
These allow us to write the stationary shock wave solution as \cite{Hossen2017aa,Karpman1975,Hasegawa1975}
\begin{eqnarray}
&&\hspace*{-1.3cm}\Psi=\Psi_m\Big[1-\tanh\bigg(\frac{\zeta}{\Delta}\bigg)\Big],
\label{1eq:39}\
\end{eqnarray}
where the amplitude $\Psi_m$ and width $\Delta$ are given by
\begin{eqnarray}
&&\hspace*{-1.3cm}\Psi_m=\frac{U_0}{A},~~~~\mbox{and}~~~~\Delta=\frac{2C}{U_0}.
\label{1eq:40}\
\end{eqnarray}
It is clear from  Eqs. \eqref{1eq:39} and \eqref{1eq:40} that the IASHWs exist,
which are formed due to the balance between nonlinearity and dissipation,
because $C>0$ and the IASHWs with $\Psi>0$ ($\Psi<0$) exist if $A>0$ ($A<0$) because $U_0>0$.
\begin{figure}[t!]
\centering
\includegraphics[width=70mm]{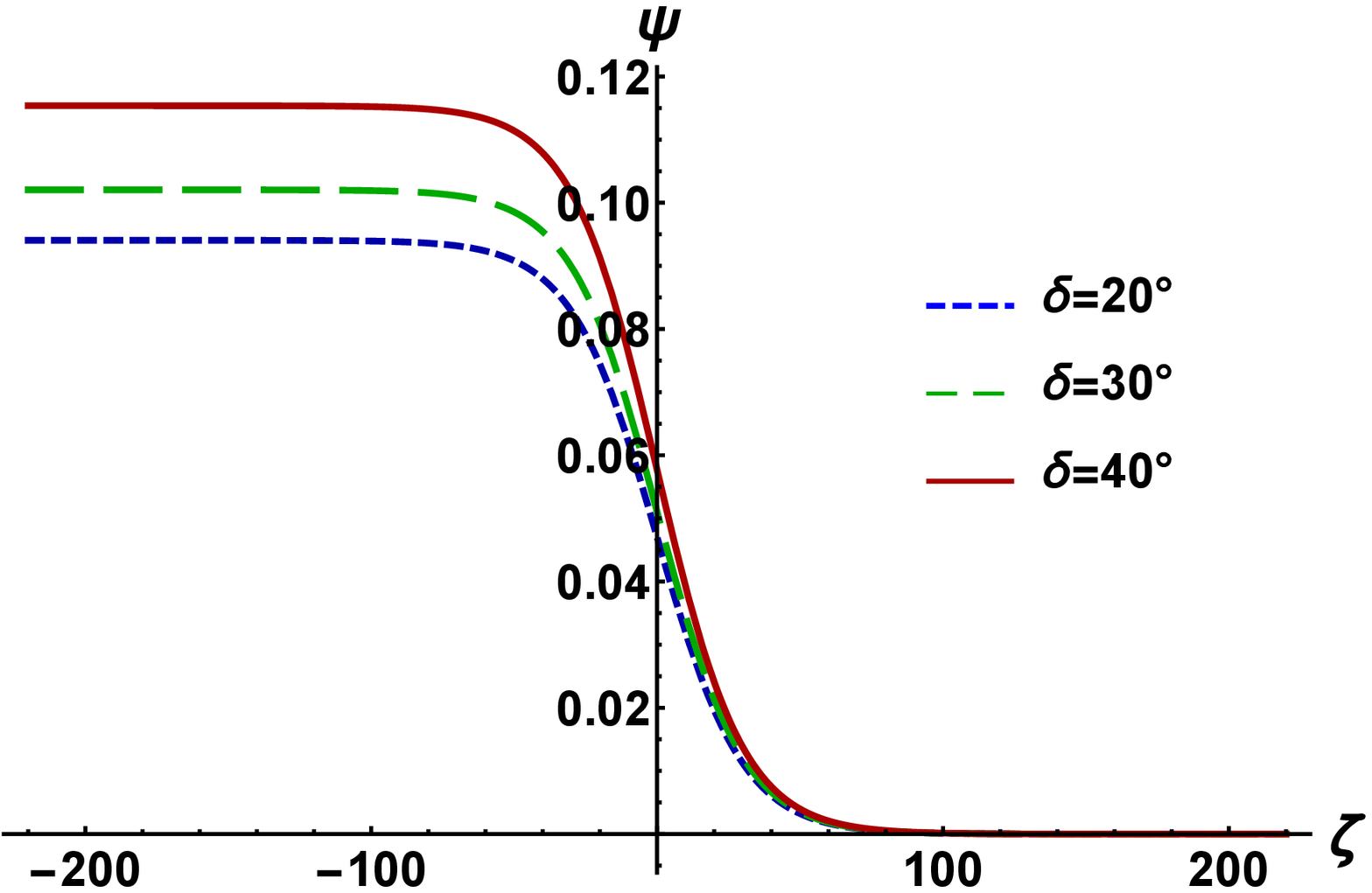}
\caption{The variation of $\Psi$ with $\zeta$ for different values of $\delta$ under the consideration $\mu_e>\mu_{ec}$.
Other plasma parameters are $\alpha_1=1.5$, $\alpha_2=0.2$, $\alpha_3=0.02$,
$\eta=0.3$, $\mu_e=0.3$, $q=1.2$, $U_0=0.01$, and $v_{p}\equiv v_{p+}$.}
\label{1Fig:F2}
\vspace{0.5cm}
\includegraphics[width=70mm]{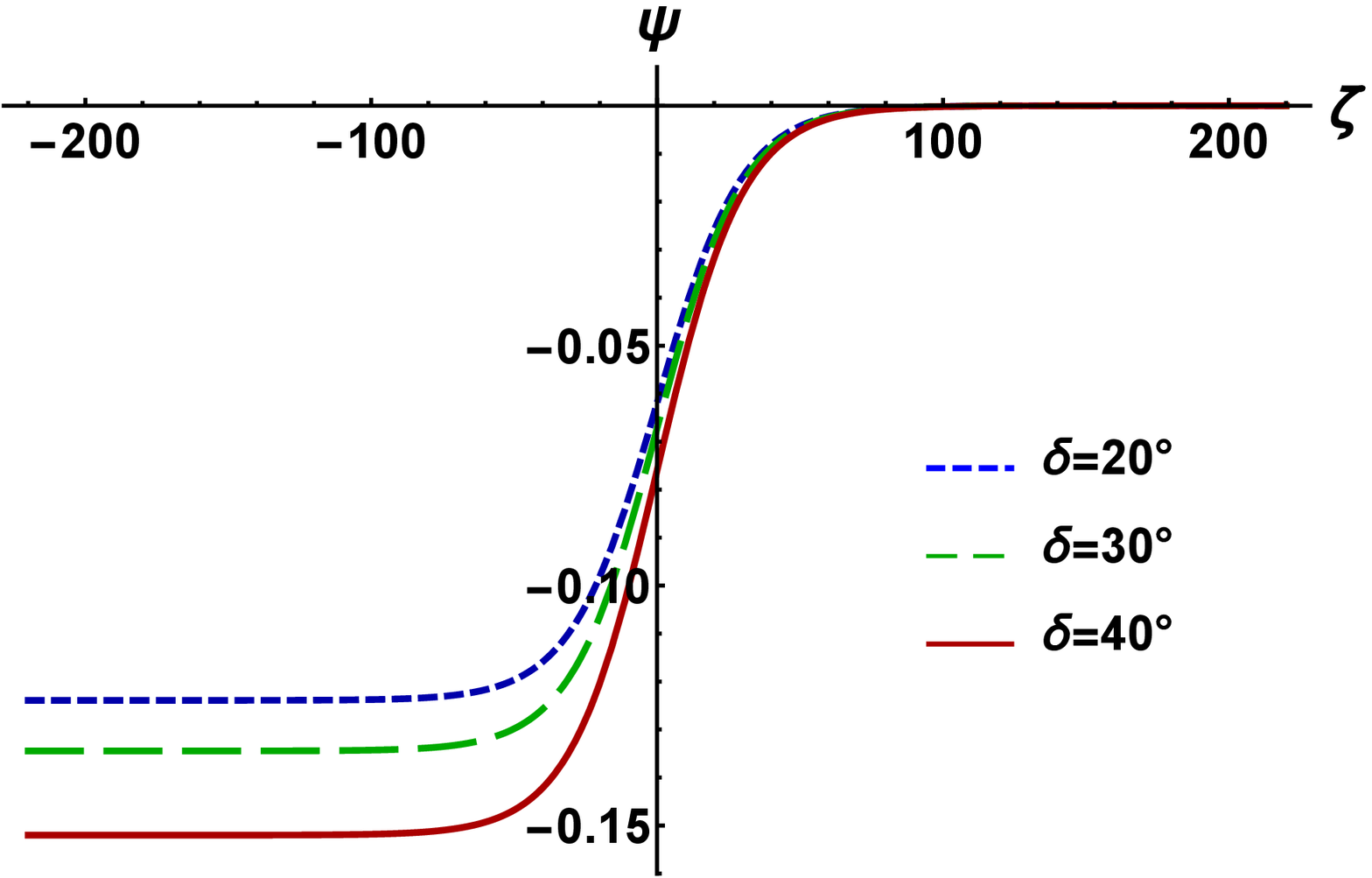}
\caption{The variation of $\Psi$ with $\zeta$ for different values of $\delta$ under the consideration $\mu_e<\mu_{ec}$.
Other plasma parameters are $\alpha_1=1.5$, $\alpha_2=0.2$, $\alpha_3=0.02$, $\eta=0.3$, $\mu_e=0.15$, $q=1.2$, $U_0=0.01$, and $v_{p}\equiv v_{p+}$.}
\label{1Fig:F3}
\end{figure}
\begin{figure}[t!]
\centering
\includegraphics[width=70mm]{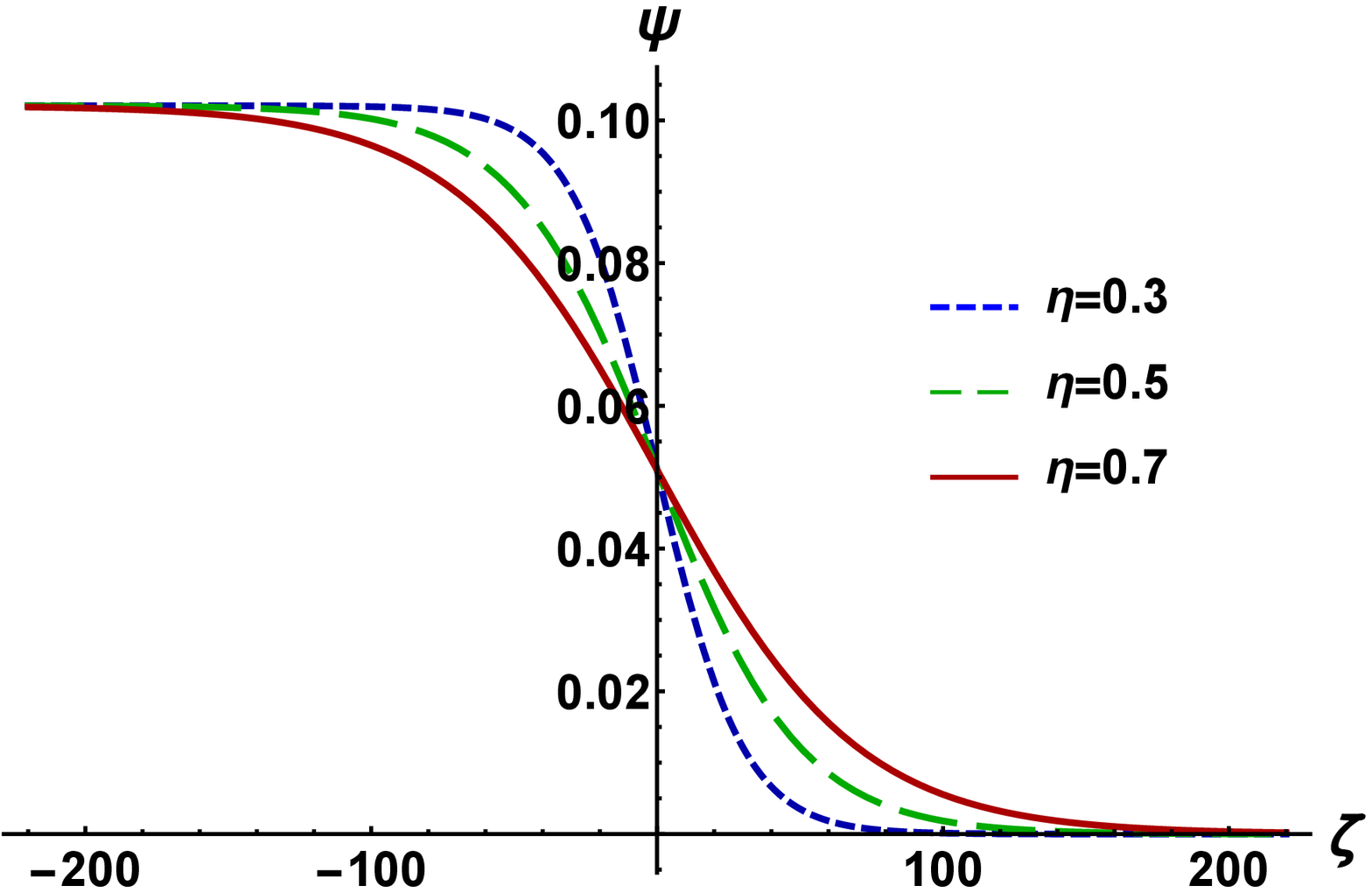}
\caption{The variation of $\Psi$ with $\zeta$ for different values of $\eta$  under the consideration $\mu_e>\mu_{ec}$.
Other plasma parameters are $\alpha_1 = 1.5$, $\alpha_2=0.2$, $\alpha_3=0.02$, $\delta=30^\circ$,
$\eta=0.3$, $\mu_e=0.3$, $q=1.2$, $U_0=0.01$, and $v_{p}\equiv v_{p+}$.}
\label{1Fig:F4}
\vspace{0.5cm}
\includegraphics[width=70mm]{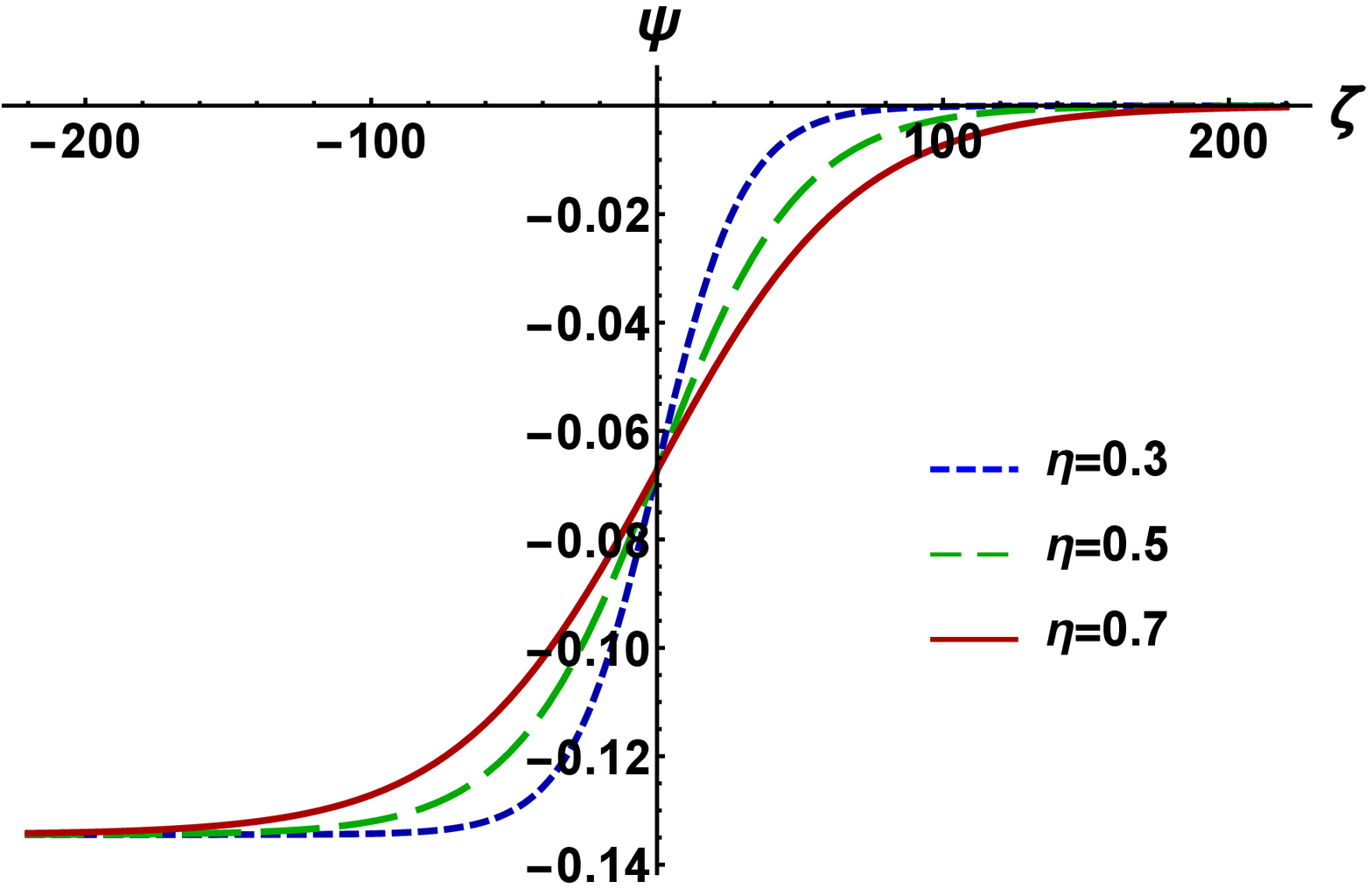}
\caption{The variation of $\Psi$ with $\zeta$ for different values of $\eta$ under the consideration $\mu_e<\mu_{ec}$.
Other plasma parameters are $\alpha_1 = 1.5$, $\alpha_2=0.2$, $\alpha_3=0.02$, $\delta=30^\circ$,
$\eta=0.3$, $\mu_e=0.15$, $q=1.2$, $U_0=0.01$, and $v_{p}\equiv v_{p+}$.}
\label{1Fig:F5}
\end{figure}
\begin{figure}[t!]
\centering
\includegraphics[width=70mm]{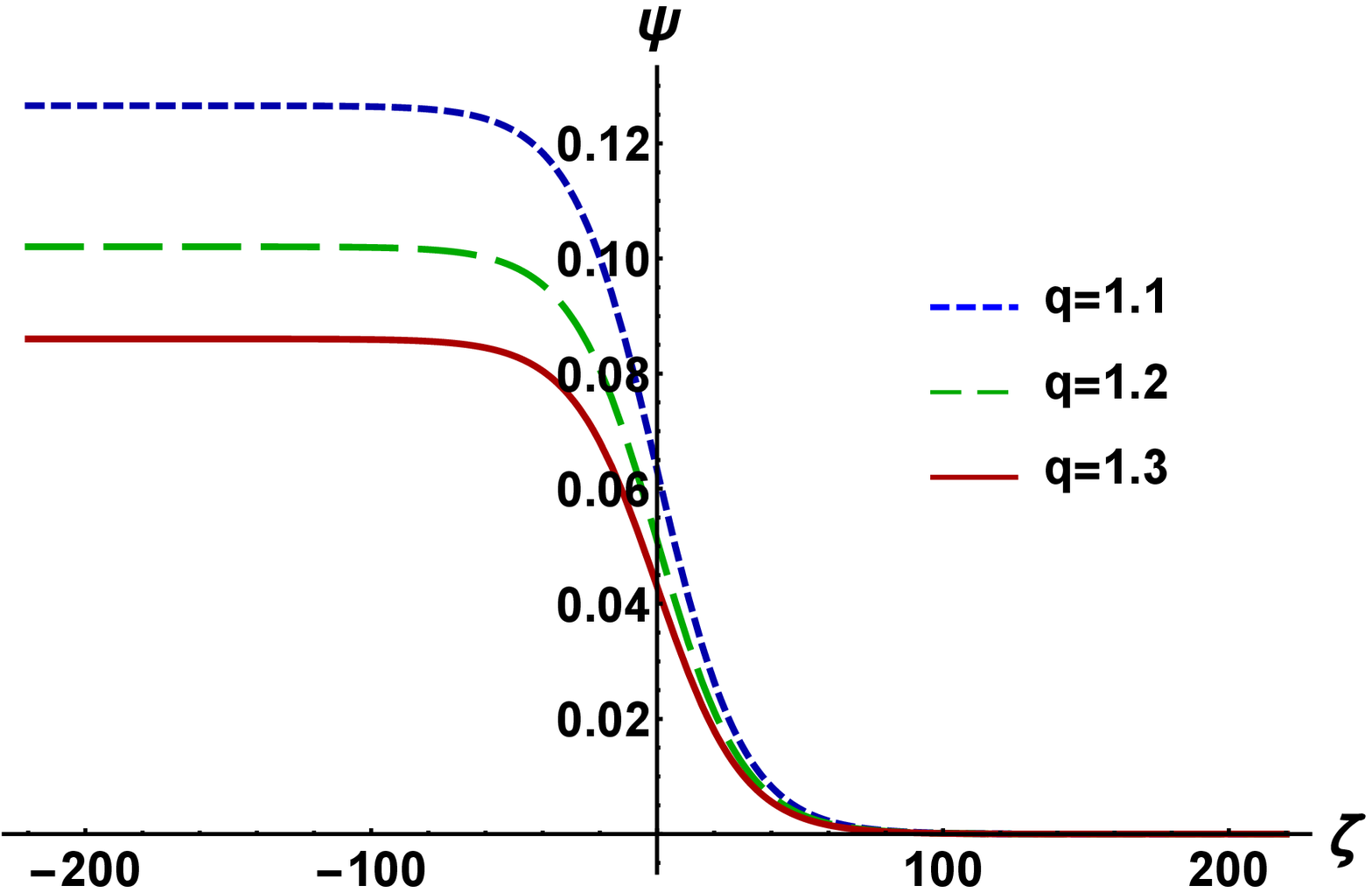}
\caption{The variation of $\Psi$ with $\zeta$ for different values of $q$  under the consideration $\mu_e>\mu_{ec}$.
Other plasma parameters are $\alpha_1 = 1.5$, $\alpha_2=0.2$, $\alpha_3=0.02$, $\delta=30^\circ$,
$\eta=0.3$, $\mu_e=0.3$, $U_0=0.01$, and $v_{p}\equiv v_{p+}$.}
\label{1Fig:F6}
\vspace{0.5cm}
\includegraphics[width=70mm]{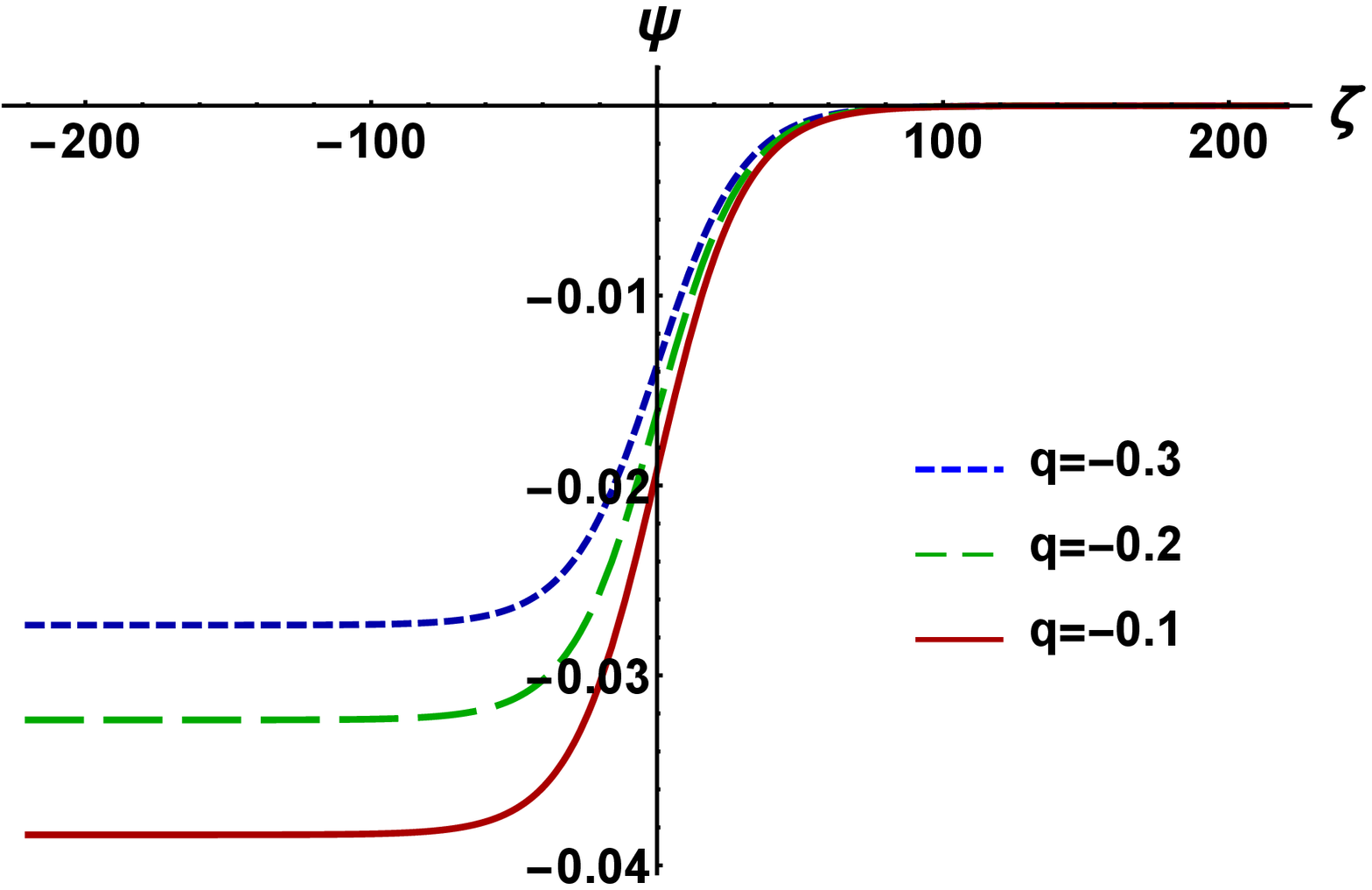}
\caption{The variation of $\Psi$ with $\zeta$ for different values of $q$ under the consideration $\mu_e>\mu_{ec}$.
Other plasma parameters are $\alpha_1 = 1.5$, $\alpha_2=0.2$, $\alpha_3=0.02$, $\delta=30^\circ$,
$\eta=0.3$, $\mu_e=0.3$, $U_0=0.01$, and $v_{p}\equiv v_{p+}$.}
\label{1Fig:F7}
\vspace{0.5cm}
\includegraphics[width=70mm]{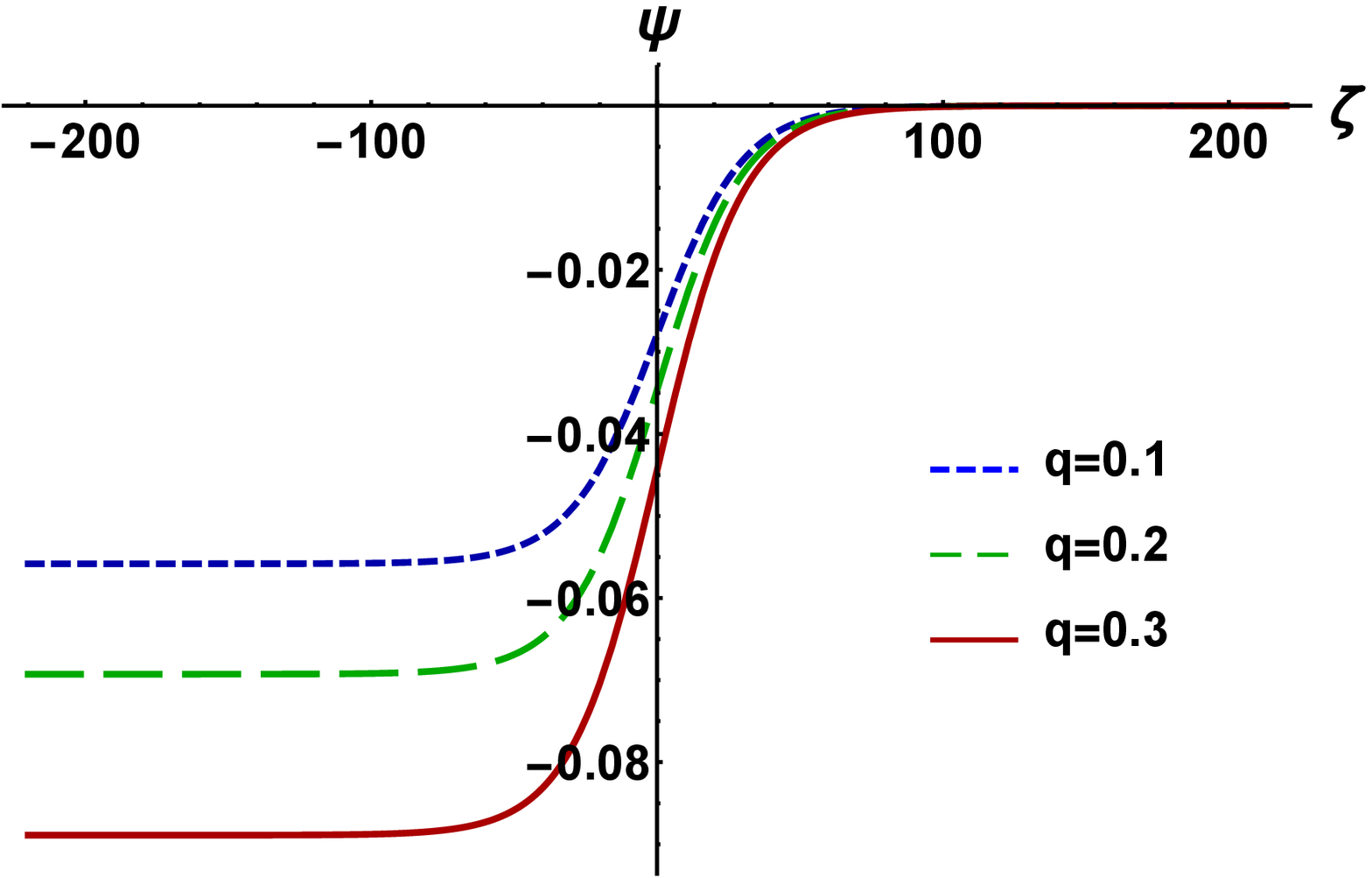}
\caption{The variation of $\Psi$ with $\zeta$ for different values of $q$ under the consideration $\mu_e>\mu_{ec}$.
Other plasma parameters are $\alpha_1 = 1.5$, $\alpha_2=0.2$, $\alpha_3=0.02$, $\delta=30^\circ$,
$\eta=0.3$, $\mu_e=0.3$, $U_0=0.01$, and $v_{p}\equiv v_{p+}$.}
\label{1Fig:F8}
\end{figure}
\begin{figure}[t!]
\centering
\includegraphics[width=70mm]{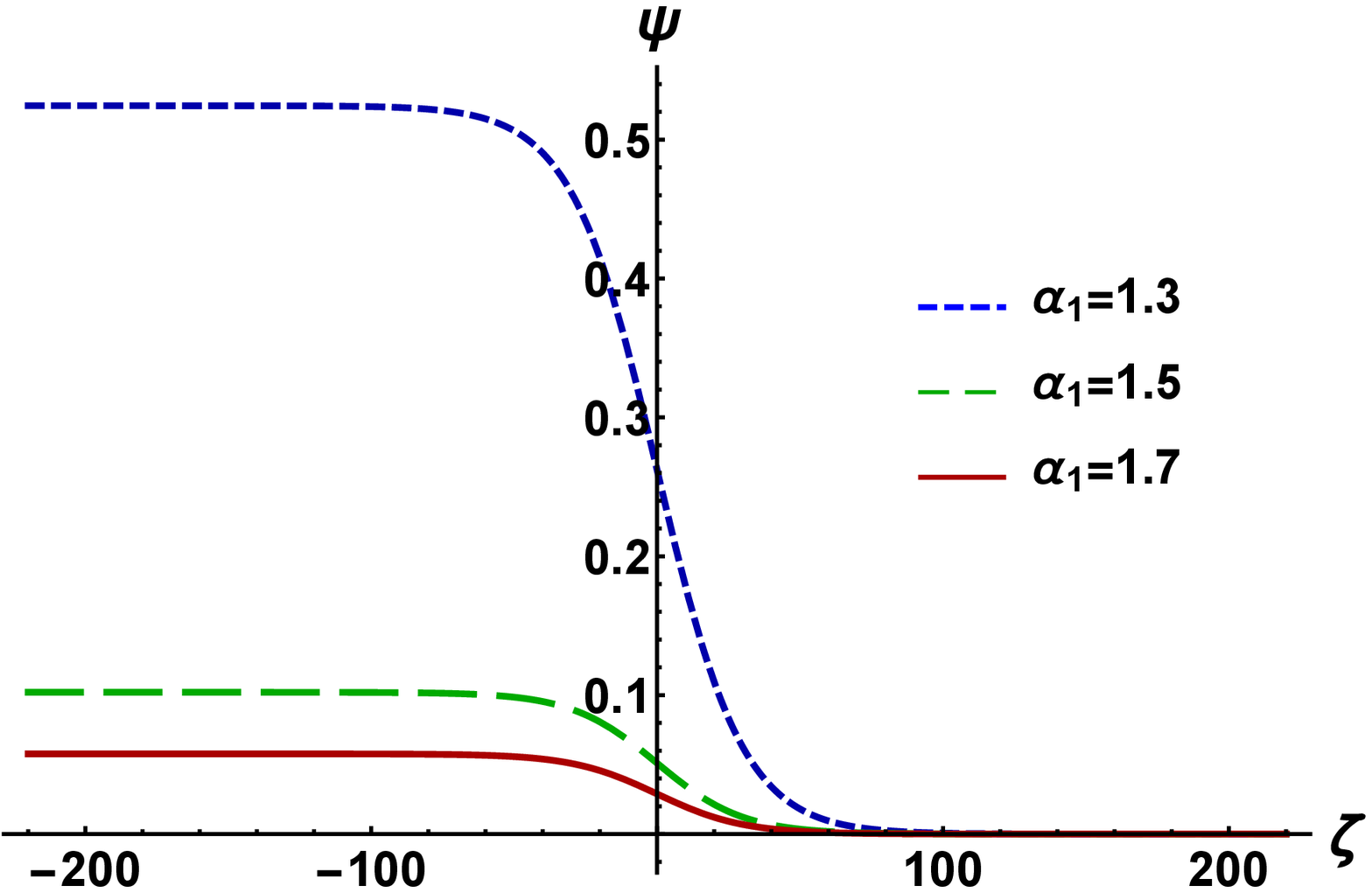}
\caption{The variation of $\Psi$ with $\zeta$ for different values of $\alpha_1$ under the consideration $\mu_e>\mu_{ec}$.
Other plasma parameters are $\alpha_2=0.2$, $\alpha_3=0.02$, $\delta=30^\circ$,
$\eta=0.3$, $\mu_e=0.3$, $q=1.2$, $U_0=0.01$, and $v_{p}\equiv v_{p+}$.}
\label{1Fig:F9}
\vspace{0.5cm}
\includegraphics[width=70mm]{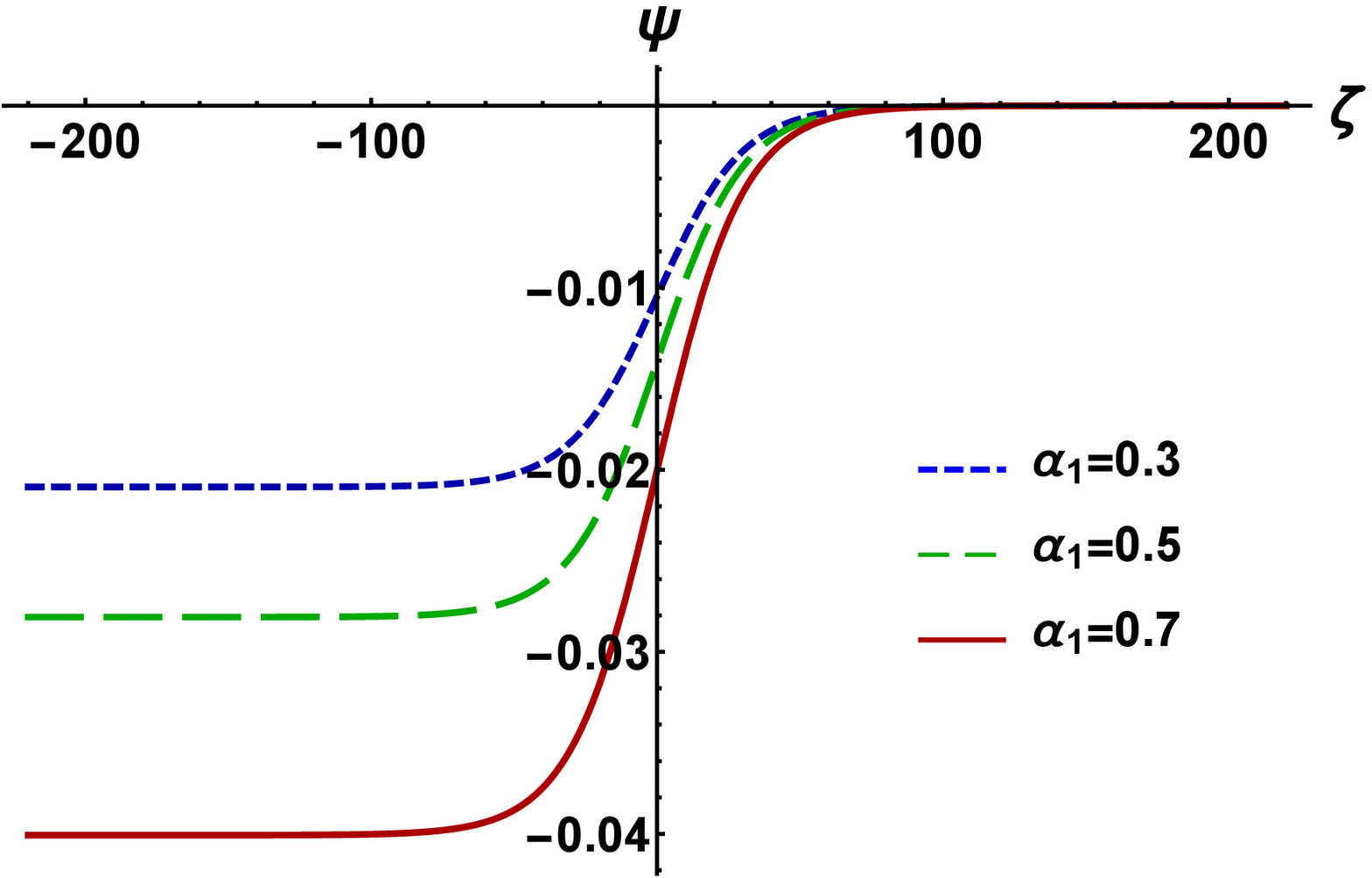}
\caption{The variation of $\Psi$ with $\zeta$ for different values of $\alpha_1$ under the consideration $\mu_e>\mu_{ec}$.
Other plasma parameters are  $\alpha_2=0.2$, $\alpha_3=0.02$, $\delta=30^\circ$,
$\eta=0.3$, $\mu_e=0.3$, $q=1.2$, $U_0=0.01$, and $v_{p}\equiv v_{p+}$.}
\label{1Fig:F10}
\end{figure}
\section{Results and discussion}
\label{1sec:Results and discussion}
The balance between nonlinearity and dissipation leads to generate IASHWs in a three-component magnetized PI plasma.
We have numerically analyzed the variation of $A$ with $\mu_e$  in the left panel of Fig. \ref{1Fig:F1:a:b},
and it is obvious from this figure that (a) $A$ can be negative, zero, and positive depending on the values of $\mu_e$;
(b) the value of $\mu_e$ for which $A$ becomes zero is known as critical value of $\mu_e$ (i.e., $\mu_{ec}$),
and the $\mu_{ec}$ for our present analysis is almost $0.2$;
and (c) the parametric regimes for the formation of positive (i.e., $\psi>0$) and negative (i.e.,  $\psi<0$) potential shock structures
can be found corresponding to $A>0$ and $A<0$. The right panel of Fig. \ref{1Fig:F1:a:b} describes the variation of
$A$ with $q$ when other plasma parameters are constant and in this case, $A$ becomes zero for the critical value of $q$ (i.e., $q=q_c\simeq0.7$).
The positive (negative) potential can exist for $q>0.7$ ($q<0.7$) [Figures are not included].

Figures \ref{1Fig:F2} and \ref{1Fig:F3} display the variation of the positive potential
shock structure under the consideration $\mu_e>\mu_{ec}$ and negative
potential shock structure under the consideration $\mu_e<\mu_{ec}$ with the oblique angle ($\delta$), respectively.
It is clear from these figures that (a) the magnitude of the amplitude of positive and
negative potential structures increases with an increase in
the value of the $\delta$, and this result agrees with the result of Hossen \textit{et al.} \cite{Hossen2017aa};
(b) the magnitude of the negative potential is always greater than the positive potential for
same plasma parameters. So, the oblique angle enhances the amplitude of the potential profiles.

Figures \ref{1Fig:F4} and \ref{1Fig:F5} illustrate the effects of the ion kinematic viscosity on
the positive (under the consideration $\mu_e>\mu_{ec}$) and negative (under the consideration $\mu_e<\mu_{ec}$) shock profiles.
It is really interesting that the magnitude of the amplitude of positive and negative shock profiles is not
effected by the variation of the ion kinematic viscosity but the steepness of the shock profile decreases with
ion kinematic viscosity, and this result agrees with the previous work of Refs. \cite{Hafez2017,Abdelwahed2016b}.

The effects of the sub-extensive electrons (i.e., $q>1$) on the positive potential
profile can be seen in Fig. \ref{1Fig:F6} under the consideration $\mu_e>\mu_{ec}$. The height of the
positive potential decreases with $q$, and this result is a good agreement with the result of
El-Labany \textit{et al.} \cite{El-Labany2020} and Hussain \textit{et al.} \cite{Hussain2013}.
Figures \ref{1Fig:F7} and \ref{1Fig:F8} illustrate the role of super-extensive electrons (i.e., $q<1$)  on the
formation of the negative potential under the consideration $\mu_e>\mu_{ec}$, and this is really
interesting that the existence of the super-extensive electron produces negative potential, and the
magnitude of the amplitude of negative potential increases with $q$. So, the orientation of the
potential profiles (positive and negative) has been organized by the sign of $q$ under the consideration $\mu_e>\mu_{ec}$.

It can be seen from the literature that the PI plasma system can support these conditions:
$m_->m_+$ (i.e., $H^+-O_2^-$ \cite{Massey1976,Sabry2009,Abdelwahed2016,Misra2009,Mushtaq2012,Jannat2015}, $Ar^+-SF_6^-$ \cite{Wong1975,Nakamura1997,Cooney1991,Nakamura1999}, and
$Xe^+-SF_6^-$ \cite{Wong1975,Nakamura1997,Cooney1991,Nakamura1999}),
$m_-=m_+$ (i.e., $H^+-H^-$ \cite{Massey1976,Sabry2009,Abdelwahed2016,Misra2009,Mushtaq2012,Jannat2015} and  $C_{60}^+-C_{60}^-$ \cite{Oohara2003,Hatakeyama2005,Oohara2005}),
and $m_-<m_+$ (i.e., $Ar^+-F^-$ \cite{Sabry2009,Abdelwahed2016}). So, in our present investigation,
we have graphically observed the variation of the electrostatic positive potential with $\alpha_1$
under the consideration of $m_->m_+$ (i.e., $\alpha_1>1$) and $\mu_e>\mu_{ec}$ in Fig. \ref{1Fig:F9}, and it is obvious from this
figure that (a) the amplitude of the positive potential decreases with an increase in the value of the negative ion mass
but increases with an increase in the value of the positive ion mass for a fixed value of their charge state;
(b) the height of the IASHWs with positive potential increases (decreases) with negative (positive) ion charge state for a constant
mass of positive and negative ion species. So, the mass and charge state of the PI play an opposite role for the formation of positive shock structure.
Figure \ref{1Fig:F10} describes the nature of the electrostatic negative potential with $\alpha_1$ under
the consideration of $m_-<m_+$ (i.e., $\alpha_1<1$) and $\mu_e>\mu_{ec}$. It is clear from this figure
that (a) due to the  $m_-<m_+$ (i.e., $\alpha_1<1$), we have observed negative potential profile even
though we have considered $\mu_e>\mu_{ec}$ (i.e., $A>0$);
(b) the existence of the heavy positive ion change the dynamics of the plasma system; and
(c) in this case, the magnitude of the amplitude of negative potential increases (decreases) with negative (positive) ion mass when other plasma parameters
are constant. So, the dynamics of the PI plasma rigourously changes with these conditions $m_->m_+$ (i.e., $\alpha_1>1$) and $m_-<m_+$ (i.e., $\alpha_1<1$).
\section{Conclusion}
\label{1sec:Conclusion}
We have studied IASHWs in a three-component magnetized PI plasma by considering kinematic viscosities
of both inertial warm positive and negative ion species, and inertialess non-extensive electrons.
The reductive perturbation method \cite{C1} is used to derive the Burgers' equation. The results that have been found from
our investigation can be summarized as follows:
\begin{itemize}
  \item The parametric regimes for the formation of positive (i.e., $\psi>0$) and negative (i.e., $\psi<0$)
  potential shock structures can be found corresponding to $A>0$ and $A<0$.
  \item The magnitude of the amplitude of positive and negative
            shock structures increases with the oblique angle ($\delta$) which arises due to the external magnetic field.
  \item The magnitude of the amplitude of positive and negative shock profiles is not effected by the variation of the ion kinematic viscosity
          but the steepness of the shock profile decreases with ion kinematic viscosity.
\end{itemize}
It may be noted here that the gravitational effect is very important
but beyond the scope of our present work. In future and for
better understanding, someone can investigate the nonlinear
propagation in a three-component PI plasma by considering the
gravitational effect. The results of our present investigation will be useful
in understanding the nonlinear phenomena both in
astrophysical environments such as upper regions of Titan's atmosphere \cite{Coates2007,Massey1976,Sabry2009,Abdelwahed2016,Misra2009,Mushtaq2012,Jannat2015,El-Labany2020},
cometary comae \cite{Chaizy1991}, ($H^+$, $O_2^-$) and ($H^+$, $H^-$) plasmas in the D and F-regions of Earth's
ionosphere \cite{Massey1976,Sabry2009,Abdelwahed2016,Misra2009,Mushtaq2012,Jannat2015}, and also
in the laboratory experiments, namely, ($Ar^+$, $F^-$) plasma \cite{Nakamura1984}, ($K^+$, $SF_6^-$) plasma \cite{Song1991,Sato1994},
neutral beam sources \cite{Bacal1979}, plasma processing reactors \cite{Gottscho1986},  ($Ar^+$, $SF_6^-$) plasma \cite{Wong1975,Nakamura1997,Cooney1991,Nakamura1999},
combustion products \cite{Sheehan1988}, plasma etching \cite{Sheehan1988}, ($Xe^+$, $F^-$) plasma \cite{Ichiki2002}, ($Ar^+$, $O_2^-$) plasma,
and Fullerene ($C_{60}^+$, $C_{60}^-$) plasma \cite{Oohara2003,Hatakeyama2005,Oohara2005}, etc.


\begin{thebibliography}{99}

\bibitem{Coates2007} A.J. Coates, \textit{et al.}, Geophys. Res. Lett. \textbf{34},
    L22103 (2007).

\bibitem{Massey1976} H. Massey, \textit{Negative Ions}, 3rd ed., (Cambridge University Press, Cambridge, 1976).

\bibitem{Sabry2009}  R. Sabry, \textit{et al.}, Phys. Plasmas \textbf{16}, 032302 (2009).

\bibitem{Abdelwahed2016} H.G. Abdelwahed, \textit{et al.}, Phys. Plasmas \textbf{23}, 022102 (2016).

\bibitem{Misra2009} A. P. Misra, Phys. Plasmas, \textbf{16}, 033702 (2009).

\bibitem{Mushtaq2012} A. Mushtaq, \textit{et al.}, Phys. Plasmas \textbf{19}, 042304 (2012).

\bibitem{Jannat2015}N. Jannat, \textit{et al.}, Commun. Theor. Phys. \textbf{64}, 479 (2015).

\bibitem{El-Labany2020} S.K. El-Labany, \textit{et al.}, Eur. Phys. J. D \textbf{74}, 104 (2020);
                         N.A. Chowdhury, \textit{et al.}, Chaos \textbf{27}, 093105 (2017); 
                         N. Ahmed, \textit{et al.}, Chaos \textbf{28}, 123107 (2018);
                         M. Hassan, \textit{et al.}, Commun. Theor. Phys. \textbf{71}, 1017 (2019);
                         S. Jahan, \textit{et al.}, Plasma Phys. Rep. \textbf{46}, 90 (2020).

\bibitem{Chaizy1991}P.H. Chaizy, \textit{et al.}, Nature (London), \textbf{349}, 393 (1991).

\bibitem{Nakamura1984} Y. Nakamura, I. Tsukabayashi, Phys. Rev. Lett. \textbf{52}, 2356 (1984).

\bibitem{Song1991} B. Song, \textit{et al.}, Phys. Fluids B \textbf{3}, 284 (1991).

\bibitem{Sato1994}N. Sato, Plasma Sources Sci. Technol. \textbf{3}, 395 (1994).

\bibitem{Bacal1979} M. Bacal, G.W. Hamilton, Phys. Rev. Lett. \textbf{42}, 1538 (1979).

\bibitem{Gottscho1986} R.A. Gottscho, C.E. Gaebe, IEEE Trans. Plasma Sci. \textbf{14}, 92 (1986).

\bibitem{Wong1975} A.Y. Wong, \textit{et al.}, Phys. Fluids \textbf{18}, 1489 (1975).

\bibitem{Nakamura1997} Y. Nakamura, \textit{et al.}, Plasma Phys. Control. Fusion \textbf{39}, 105 (1997).

\bibitem{Cooney1991}J.L. Cooney, \textit{et al.}, Phys. Fluids B \textbf{3}, 2758 (1991).

\bibitem{Nakamura1999}Y. Nakamura, \textit{et al.}, Phys. Plasmas \textbf{6}, 3466 (1999).

\bibitem{Sheehan1988} D.P. Sheehan, N. Rynn, Rev. Sci. lnstrum. \textbf{59}, 8 (1988).

\bibitem{Ichiki2002} R. Ichiki, \textit{et al.}, Phys. Plasmas \textbf{9}, 4481 (2002).

\bibitem{Oohara2003} W. Oohara, R. Hatakeyama, Phys. Rev. Lett. \textbf{91}, 205005 (2003).

\bibitem{Hatakeyama2005} R. Hatakeyama, W. Oohara, Phys. Scripta \textbf{116}, 101 (2005).

\bibitem{Oohara2005} W. Oohara, \textit{et al.}, Phys. Rev. Lett. \textbf{95}, 175003 (2005).

\bibitem{Hansen2005} S.H. Hansen, New Astron.   \textbf{10}, 371 (2005).

\bibitem{Asbridge1968} J.R. Asbridge,  \textit{et al.}, J. Geophys. Res. \textbf{73}, 5777 (1968).

\bibitem{Lundlin1989}R. Lundlin, \textit{et al.},  Nature (London) \textbf{341}, 609 (1989).

\bibitem{Futaana2003} Y. Futaana, \textit{et al.},  J. Geophys. Res. \textbf{108}, 1025 (2003).

\bibitem{Krimigis1983} S.M. Krimigis, \textit{et al.}, J. Geophys. Res. \textbf{88}, 8871 (1983).

\bibitem{Renyi1955} A. R\'{e}nyi, Acta Math. Acad. Sci. Hung. \textbf{6}, 285 (1955).

\bibitem{Tsallis1988} C. Tsallis, J. Stat. Phys. \textbf{52}, 479 (1988).

\bibitem{Hussain2013} S. Hussain, \textit{et al.}, Phys. Plasmas \textbf{20}, 092303 (2013).

\bibitem{Tribeche2010} M. Tribeche, L. Djebarni, R. Amour, Phys. Plasmas \textbf{17}, 042114 (2010).

\bibitem{Hafez2017} M.G. Hafez, \textit{et al.}, Plasma Phys. Rep. \textbf{43}, 499 (2017).

\bibitem{Abdelwahed2016b} H.G. Abdelwahed, \textit{et al.}, J. Exp. Theor. Phys. \textbf{122}, 1111 (2016).

\bibitem{Hossen2017aa} M.M. Hossen, \textit{et al.}, High Energy Density Phys. \textbf{24}, 9 (2017).

\bibitem{Atteyaa2018} A. Atteya, S. Sultana, R. Schlickeiser, Chin. J. Phys. \textbf{56}, 1931 (2018).

\bibitem{Adhikary2012} N.C. Adhikary, Phys. Lett. A \textbf{376}, 1460 (2012).

\bibitem{Dev2018}A.N. Dev, M.K. Deka, Phys. Plasmas \textbf{25}, 072117 (2018).

\bibitem{Dev2016} A.N. Dev, \textit{et al.}, Chin. Phys. B \textbf{25}, 105202 (2016).

\bibitem{Dev2014}A.N. Dev, \textit{et al.}, Commun. Theor. Phys. \textbf{62}, 875 (2014).

\bibitem{Deka2018}M.K. Deka, A.N. Dev,  Plasma Phys. Rep. \textbf{44}, 965 (2018).

\bibitem{Sahu2014} B. Sahu, A. Sinha, R. Roychoudhury, Phys. Plasmas \textbf{21}, 103701 (2014).

\bibitem{Washimi1966} H. Washimi, T. Taniuti, Phys. Rev. Lett. \textbf{17}, 996 (1966).

\bibitem{Karpman1975} V.I. Karpman, \textit{Nonlinear Waves in Dispersive Media}, (Pergamon Press, Oxford, 1975).

\bibitem{Hasegawa1975} A. Hasegawa, \textit{Plasma Instabilities and Nonlinear Effects}, (Springer-Verlag, Berlin, 1975).

\bibitem{C1} M.H. Rahman,\textit{et al.}, Phys. Plasmas \textbf{25}, 102118 (2018);
             N.A. Chowdhury, \textit{et al.}, Phys. plasmas \textbf{24}, 113701 (2017);
             M.H. Rahman, \textit{et al.}, Chin. J. Phys. \textbf{56}, 2061 (2018);
             N.A. Chowdhury, \textit{et al.}, Vacuum \textbf{147}, 31 (2018);
             R.K. Shikha, \textit{et al.}, Eur. Phys. J. D \textbf{73}, 177 (2019);
             N.A. Chowdhury, \textit{et al.}, Contrib. Plasma Phys. \textbf{58}, 870 (2018);
             N.A. Chowdhury, \textit{et al.}, Plasma Phys. Rep. \textbf{45}, 459 (2019);
             S.K. Paul, \textit{et al.}, Pramana J. Phys \textbf{94}, 58 (2020);
             T.I. Rajib, \textit{et al.}, Phys. plasmas \textbf{26}, 123701 (2019);
             S. Jahan, \textit{et al.}, Commun. Theor. Phys. \textbf{71}, 327 (2019).

\end{thebibliography}
\end{document}